\documentclass{PoS}
\usepackage{amsmath,amsfonts}
\def \beq  {\begin{equation}}
\def \eeq  {\end{equation}}
\def \beqar {\begin{eqnarray}}
\def \eeqar {\end{eqnarray}}
\def\sqr#1#2{{\vcenter{\vbox{\hrule height.#2pt
\hbox{\vrule width.#2pt height#1pt \kern#1pt
\vrule width.#2pt}\hrule height.#2pt}}}}

\def\la {{\langle}}
\def\ra {{\rangle}}
\def\vx {{\vec x}}
\def\vy {{\vec y}}
\def\vk {{\vec k}}
\def\vf {{\varphi}}

\def\Tr {{\rm Tr}}

\def\bD {\bar{D}}

\def\vk {\vec{k}}
\def\vx {{\vec x}}
\def\vz {\vec{z}}
\def\vy{\vec{y}}

\def\vw {\vec{w}}

\def\del {\partial}
\def\bdel{\bar{\partial}}

\def\d {\delta}

\def\bz {{\bar{z}}}

\def\vf {{\varphi}}

\def\half{\textstyle{1\over 2}}

\def \PL {{Phys. Lett.}}

\def \NP {{Nucl. Phys.}}

\def \PR {{Phys. Rev.}}

\title{Gauge-invariant Mass Terms and Wave Functions}

\ShortTitle{Yang-Mills (2+1)}

\author{\speaker{V.P. NAIR}\thanks{Supported in part by NSF grant PHY-0855515 and a PSC-CUNY grant}\\
        Physics Department, City College of the CUNY, New York, NY 10031\\
        E-mail: \email{vpn@sci.ccny.cuny.edu}}

\abstract{We outline a method of relating the quantum effective action and the ground state wave function of a field theory. This method, along with a gauge-invariant
mass term and the previously obtained vacuum wave function, is used to arrive at the leading terms of the 3d-covariant quantum effective action for the Yang-Mills theory in three (or 2+1) dimensions. Some features of this effective action are also discussed.}

\FullConference{International Workshop on QCD Green's Functions, Confinement and Phenomenology,\\
		September 05-09, 2011\\
		Trento, Italy}
\begin{document}

\section{Introduction}

As many of you know, my collaborators and I have worked for many years on developing a Hamiltonian approach to the Yang-Mills theory in three (or $2+1$) dimensions, which has, among other things, led to a vacuum wave function \cite{KN1}-\cite{KKN2}. This has, in turn, led to an analytic formula for the string tension, which is in good agreement with the lattice estimates \cite{KKN2}. A Hamiltonian analysis, by its very nature, is not manifestly covariant. While it is not necessary to have {\it manifest} covariance,
there is still a definite advantage and simplification in a manifestly covariant formalism, such as 
what we would have using quantum effective actions.
So, in this talk,
I would like to discuss the question of whether we can find a quantum effective action
for Yang-Mills ($2+1$) which is manifestly covariant.
Starting from the Yang-Mills action, we can quantize in the Hamiltonian formalism and find the vacuum wave function by solving the Schr\"odinger equation. In principle, a direct evaluation
of the functional integral of the theory (which cannot be done in practice) would lead to the quantum effective action or the generating functional for the one-article irreducible vertices.
The question we want to ask is: Given the vacuum wave function, can we get the quantum effective action without having to go through the functional integration, in other words, can we directly relate the wave function and the quantum effective action?
We will give an explicit formula for relating the vacuum wave function of any field theory to the quantum effective action, and then applying it to the Yang-Mills case, we will obtain an effective action which is manifestly covariant.

In order to frame the problem in a quantitive way, we will need some features of our previous work
\cite{KN1}-\cite{KKN2}.
Our
Hamiltonian analysis was done in the $A_0 =0 $ gauge, with the spatial components of the gauge potentials represented as
\beq
A_z = {\half} (A_1 + i A_2) = - \del M ~M^{-1}, \hskip .2in A_{\bz} = {\half} (A_1 - i A_2 )
= M^{\dagger -1} \bdel M^\dagger
\label{1}
\eeq
where $z = x_1 - i x_2$, $\bz = x_1 + i x_2$ are complex combinations of the coordinates.
$M$ is an element of the complexified group; i.e., it is an $SL(N, {\mathbb{C}})$-matrix
if the gauge transformations take values in $SU(N)$.
Imposing the Gauss law is equivalent to requiring that the wave functions be gauge-invariant.
They are  functions of $H = M^\dagger M$, with the inner product
\beq
\la 1\vert 2\ra = \int d\mu (H) \exp [2~c_A~S_{wzw}(H)]~ \Psi_1^*\, \Psi_2
\label{2}
\eeq
The hermitian gauge-invariant matrix $H$ parametrizes
$SL(N, {\mathbb{C}})/SU(N)$ and $d\mu (H)$ is the Haar measure for $H$.
Also, in (\ref{2}), 
$S_{wzw}$ is the Wess-Zumino-Witten action given by
\beq
S_{wzw} (H) = {1 \over {2 \pi}} \int \Tr (\partial H ~\bdel
H^{-1}) +{i \over {12 \pi}} \int \epsilon ^{\mu \nu \alpha} \Tr (
H^{-1}
\partial _{\mu} H~ H^{-1}
\partial _{\nu}H ~H^{-1} \partial _{\alpha}H)
\label{3}
\eeq
Further, $c_A$ is the adjoint Casimir value, equal to $N$ for $SU(N)$.

All observables, including the Hamiltonian, are functions of the current $J$
of the WZW action, which, up to a proportionality factor, is given by
\beq
J = {2 \over e} \, \del H ~ H^{-1}
\label{4}
\eeq
In terms of $J$, we have
${\cal H} = {\cal H}_0 + {\cal H}_1$, where
\beqar
{\cal H}_0 &=& m  \int_z J_a (\vz) {\d \over {\d J_a (\vz)}} + {2\over \pi} \int _{z,w} 
 {1\over (z-w)^2} {\d \over {\d J_a (\vw)}} {\d \over {\d
J_a (\vz)}}\nonumber\\
&&\hskip .4in + {1\over 2} \int_x :\bdel J^a(x) ~\bdel J^a(x):\label{5}\\
{\cal H}_1& =&  i ~{e}~ f_{abc} \int_{z,w}  {J^c(\vw) \over \pi (z-w)}~ {\d \over {\d J_a (\vw)}} {\d \over {\d
J_b (\vz)}} \nonumber
\eeqar
where $m = e^2 c_A /2\pi$.
We have displayed the Hamiltonian in a way that anticipates our method of solving the 
Schr\"odinger equation.
We keep all terms in ${\cal H}_0$
at the lowest  order and treat ${\cal H}_1$ as a perturbation. 
Since $m = e^2 c_A/2\pi$, this is not standard perturbation theory. In fact, in the latter, one would expand in powers of $m$ as well. 
Our expansion corresponds to a partially resummed version, with certain particular infinite sequences of perturbative terms being summed up \cite{KNY}.
To keep track of various orders without confusion, starting from
(\ref{5}), we consider
$m$ and $e$ as independent parameters, only setting $m = e^2 c_A/2\pi$ at the end.
The lowest order computation of the wave function in this scheme is the one presented in \cite{KKN2}; it led to the formula for the string tension as $\sigma_R = e^4 c_A c_R /4 \pi$. 
More recently, we have calculated corrections to this formula, carrying out the expansion to the next higher order (which still involves an infinity of terms); the final value of the correction to the string tension was to be small, of the order of 
$-0.03$ to $-2.8$ percent \cite{KNY}.

The measure of integration in (\ref{2}) will play an important role in what follows. To illustrate this,
we will need a short argument from \cite{KNrobust} on the nature of the wave function.
For this, we first absorb the factor $e^{2 c_A S_{wzw}}$ from the measure of integration
 in (\ref{2}) into the wave function by writing $\Psi = e^{- c_A S_{wzw}} \, \Phi$. 
 For the action on $\Phi$, the Hamiltonian operator is given by
${\cal H} \rightarrow e^{- c_A S_{wzw}}\, {\cal H}\, e^{- c_A S_{wzw}}$.
The matrix $H$ is then expanded as  $H = \exp( t_a \vf^a ) \, \approx  1 + t_a \vf^a + \cdots$; this ``small $\vf$" expansion is appropriate for a (resummed) perturbation theory. 
It is easily verified that the Hamiltonian then takes the form
\beq
{\cal H}= {1\over 2}\int \left[ -{\delta^2 \over \delta \phi^2} +\phi (-\nabla^2 +m^2)\phi
+\cdots\right] \label{6}
\eeq
where $\phi _a (\vk) = \sqrt {{c_A k \bar{k} }/ (2 \pi m)}~~ \vf _a (\vk)$. This is the Hamiltonian for a field of mass $m$. The ground state (or vacuum) wave function for this is easily obtained by solving
${\cal H} \, \Phi_0 = 0$, and it is given by
\beq
\Phi_0 \approx \exp \left[ - {1\over 2} \int \phi^a \sqrt{ m^2 - \nabla^2} ~\phi^a \right]
\label{7}
\eeq
Transforming back to the $\Psi$'s, we find
\beq
\Psi_0 \approx \exp \left[ - {c_A \over \pi m} \int (\bdel \del \vf^a) \left[{1\over  \sqrt{- \nabla^2 + m^2}~ + m}\right]
(\bdel \del \vf^a) + \cdots \right] \label{8}
\eeq
The next key step is as follows.
On general grounds, we know that the  full wave function must be a functional of the current $J$
\cite{{KKN1},{KNrobust}}. So we can ask: Is there a functional of the current $J$ which reduces to
(\ref{8}) in the small $\vf$ approximation, when $J^a \approx (2/ e) \del \vf ^a+{\cal O}(\vf^2)$?
The only  answer for this is
\beq
\Psi_0  =  \exp \left[ - {2\pi^2 \over e^2 c_A^2} \int \bdel J^a (x) \left[{1\over  \sqrt{- \nabla^2 + m^2}~ + m}\right]_{x,y} ~\bdel J^a (y) ~+\cdots\right]
\label{9}
\eeq
We arrived at this without directly solving the Schr\"odinger equation for
(\ref{5}), even though we did solve for the ground state of (\ref{6}) in the $\Phi$-description.
But the result of solving directly using (\ref{5}), which is given in \cite{KKN2},  is identical.

The denominator of the integral kernel appearing in the exponent in
(\ref{9}) can be rationalized to rewrite this wave function as
\beq
\Psi_0  =  \exp \left[ - {2\pi^2 \over e^2 c_A^2} \int \bdel J^a (x) \left[{\sqrt{k^2 + m^2}~  { {- m}} \over k^2}\right]_{x,y} ~\bdel J^a (y) ~+\cdots\right]
\label{10}
\eeq
In the integral kernel, the term $\sqrt{k^2 + m^2} \, / k^2 $ is due to the fact that we have a mass for the fields $\phi$, while the second part $- m /k^2$ is from transforming using $e^{c_A S_{wzw}}$ from the measure. We have presented this argument for the wave function to emphasize that the
measure of integration is crucial in both generating the mass term and also in providing the
$-m/k^2$ term.
The kernel should have the $-m/k^2$ term to give
the low momentum limit
\beq
{ \sqrt{k^2 + m^2}\, - m \over k^2} \approx {1\over 2 m},
\label{11}
\eeq
which means that the exponent in $\Psi_0^* \Psi_0$ can be approximated, for long wave length modes, by the two-dimensional Yang-Mills action,
$\int \bdel J \bdel J \sim \int F^2 / 4 g^2$, $g^2 = m e^2$.
This was, in turn, the key to obtaining the formula for the string tension.

If we write the wave function (\ref{10}) in terms of the usual gauge potentials $A$, then
\beq
\Psi_0 \approx  \exp \left[ - {1\over 2} \int A^{aT}_i (x) \left[{\sqrt{k^2 + m^2}~  {{- m}} }\right]_{x,y} ~A^{aT}_i (y) ~+\cdots\right]
\label{12}
\eeq
where we have used the transverse component of $A_i^a$ as the gauge-invariant variable.
To the quadratic order that we are interested in, this formula (\ref{12}) is adequate.

The structure of this wave function also answers another interesting question. If we have a mass for the gluon fields, we would expect the interactions mediated by gluons to be of short range.
Yet, an area law for the Wilson loop shows there are long range interactions, approximately
described as a linear potential for long distances. How can we have compatibility between these two? The wave function has part which essentially represents the massive gluon, but the
$-m$-term in the kernel shows that long range potentials are possible.

The basic question we posed earlier can be rephrased as follows:
Can we find a covariant three-dimensional quantum
effective action which will give this wave function (\ref{12}) including the crucial $- m /k^2$ term in the kernel?

\section{Relating wave functions and the effective action}

The wave function (\ref{12}) shows that part of what we need is mass for the gluon, which, of course, must be gauge-invariant.
But such a mass term must be nonlocal, including nonlocality in time;
so is an effective action, in general.
A Hamiltonian analysis for a nonlocal action is difficult. In any case, we are not supposed to set up a Hamiltonian for the effective action, it is to be set up for the original Yang-Mills action.
So what we need is a more direct way to relate the quantum effective action and wave functions.
This can be done in series of five steps as follows.
We will use a scalar field to illustrate this basic connection.
\vskip .05in\noindent
\underline{Step 1}:
\vskip .05in\noindent
Let $\{ \, \vert \alpha \ra \,\}$ denote a complete set of energy eigenstates. Then
\beqar
\la \vf \vert e^{-  \beta \, {\cal H}} \vert \vf' \ra &=&
\sum_\alpha \la \vf \vert \alpha \ra \, \la \alpha \vert \vf'\ra\, e^{-\beta E_\alpha}
=
\sum_\alpha \Psi_\alpha (\vf) \, \Psi^*_\alpha (\vf') \, e^{- \beta E_\alpha}\nonumber\\
&\rightarrow& \Psi_0 (\vf ) \, \Psi^*_0 (\vf') \, e^{-\beta E_0},\hskip .3in {\rm as}~~ {\beta \rightarrow \infty}
\label{13}
\eeqar
This shows that we can obtain the ground state wave function
$\Psi_0 (\vf )$ by calculating this matrix element with fixed boundary values of the field at the Euclidean time-boundaries, $\tau =0, \, \beta$.
\vskip .05in\noindent
\underline{Step 2}:
\vskip .05in\noindent
Our next step is to express this matrix element as a functional integral,
\beq
\la \vf \vert e^{-  \beta \, {\cal H}} \vert \vf' \ra  = \int [d\phi]\, e^{- S(\phi)} 
= \int [d\eta] \, e^{- S ( \chi + \eta )}
\label{14}
\eeq
We have written $\phi$ as $\phi = \chi + \eta$, where the boundary conditions
are
\begin{align}
\chi (0 , \vx ) &= \vf' (\vx ), \hskip .3in \chi (\beta, \vx ) = \vf (\vx )\nonumber\\
\eta (0, \vx ) &= \eta (\beta, \vx )  = 0
\label{15}
\end{align}
$\chi (\tau, \vx )$ is taken to be a fixed field configuration with the boundary values specified; it contains no additional degree of freedom to be integrated in (\ref{14}). Since $\chi$ has the 
correct boundary values, $\eta$ must vanish at both $\tau =0$ and $\tau = \beta$. Thus, in carrying out the $\eta$-integration in (\ref{14}), Dirichlet conditions in $\tau$ must be used for the $\eta$-propagator. 
\vskip .05in\noindent
\underline{Step 3}:
\vskip .05in\noindent
Rather than explicitly carrying out the $\eta$-integration, the result can be expressed in terms of the quantum effective action $\Gamma [\chi ]$.
We recall that $\Gamma [\chi ]$is defined, {\it for arbitray} $\chi$,
by
\beq
e^{- \Gamma (\chi)} = \int [d\eta] \, \exp\left[ - S(\chi + \eta ) + \int {\delta \Gamma \over \delta \chi}
\,\eta \right]
\label{16}
\eeq
This equation shows that, if we choose $\chi$ as a solution of $\delta \Gamma /\delta \chi \,
= 0$, with the boundary behavior $\chi \rightarrow \vf'$ at $\tau =0$ and
$\chi \rightarrow \vf $ at $\tau = \beta$, and with $\eta$ going to zero at both ends,
then
\beqar
e^{- \Gamma } &=& \int [d\eta]\, e^{- S (\chi +\eta )} = \la \vf\vert e^{-\beta {\cal H}} \vert \vf'\ra\nonumber\\
&\rightarrow& \Psi_0 (\vf) \Psi^*_0 (\vf') \, e^{-\beta E_0}, \hskip .2in {\rm as}~ \beta \rightarrow \infty
\label{17}
\eeqar
where we have used (\ref{14}) for the second equality.
The procedure to get the wave function from the effective action is thus the following.
\begin{quotation}
\noindent Solve the equation
\beq
{\delta \Gamma \over \delta \chi} = 0
\label{18}
\eeq
for $\chi$, subject to the boundary conditions (\ref{15}), and substitute the solution back in 
$\Gamma (\chi )$.
Then $e^{-\Gamma(\chi)}$, which is now a functional of $\vf'$, $\vf$, will give
 $\Psi_0 (\vf )$ as $\beta$ becomes large. 
 \end{quotation}
In practice, the evaluation of $\Gamma$ at its critical point can be simplified a bit further, at least for the case of interest to us, as follows.
\vskip .05in\noindent
\underline{Step 4}:
\vskip .05in\noindent
We will denote $\Gamma$ evaluated on the solution $\chi_*$ of (\ref{18}),
 subject to the boundary values (\ref{15}), by $W$.
Now, if we vary the boundary value $\vf$ of $\chi$, and also change $\beta$ slightly, the resulting variation of $\Gamma$ or $W$ can be brought to the form
\beq
\delta W = \delta \Gamma[\chi_*] = \int d^2x~ \Pi \, \delta \vf ~+~ {\cal H}_E \, \delta \beta
\label{19}
\eeq
(This is clear since $\delta W$ must be linear in the variations $\delta \vf$ and
$\delta \beta$.) Equation (\ref{19}) defines $\Pi$ (which may depend on the time-derivatives of $\vf$) and also the Euclidean Hamiltonian ${\cal H}_E$. Generally, ${\cal H}_E$ is not positive semi-definite.
In principle, we can have terms involving $\delta \chi$ which are three-volume integrals, but
since we are evaluating $\Gamma$ on the solution of (\ref{18}), 
such terms are zero.
\vskip .05in\noindent
\underline{Step 5}:
\vskip .05in\noindent
Generally, ${\cal H}_E$ can have a contribution corresponding to the zero-point energy,
but for a relativistically invariant vacuum, we know that the zero-point energy  must be zero.
\footnote{This is well known and can be easily seen from taking the vacuum expectation value of the commutation rule
$[P_i , K_j ] = i \delta_{ij}\, {\cal H}$ where $P_i$ is the momentum operator and $K_j$ is the Lorentz boost generator. For a Lorentz-invariant vacuum, i.e., with no spontaneous breaking of Lorentz symmetry, we get $\la 0 \vert {\cal H}\vert 0\ra =0$. Thus any regularization used for
the Hamiltonian or the effective action must satisfy the condition of vanishing zero-point energy.}
Therefore, for the ground state or vacuum wave function, we can impose ${\cal H}_E = 0$. Further, from (\ref{19}), $\Pi$ may be taken as $\delta W / \delta \vf$.
Thus we can find $W$ by solving the equations
\beq
{\cal H} _E = 0, \hskip .3in \Pi = {\delta W \over \delta \vf}
\label{20}
\eeq
The ground state wave function is then given by $ \Psi_0 = e^{-W}$. 
This last step of solving (\ref{20}) is obviously
a Euclidean version of the usual Hamilton-Jacobi approach.

\section{The effective action for Yang-Mills (2+1)}

We are now in a position to state our main result. The leading terms of the quantum effective
action for 3-dimensional Yang-Mills theory are given by
\beq
\Gamma = \int {1\over 4} F^a_{\mu\nu} F^a_{\mu\nu} + S_{m}(A) 
+ (\sigma^\mu D_\mu \Phi_A)^{a\dagger} (\sigma^\nu D_\nu \Phi_A)^{a} ~+~ \cdots
\label{21}
\eeq
where $S_m(A)$ is a gauge-invariant nonlocal mass term for the gauge field.
The particular choice of this mass term is not important at this stage, we will comment on this later.
$\Phi^a_A$, $a = 1, 2, \cdots, (N^2 -1)$,  $A= 1, 2$, is a complex field transforming according to the adjoint representation of $SU(N)$, and transforming as a 2-component spinor under the Lorentz group.
$\sigma^\mu$, $\mu = 1, 2, 3$, are the Pauli matrices and $D_\mu$ denotes the gauge-covariant derivative. 
A complex spinor field with a quadratic derivative term in the action may raise worries about the spin-statistics connection, but , for the action (\ref{21}), $\Phi_A^a$ is not to be considered as an observable field. It is to be viewed simply as an auxiliary field used to represent a nonlocal term in $\Gamma$ and used to capture the 
physics of the wave function (\ref{10}) or (\ref{12}). The action (\ref{21}) has an additional $U(1)$ symmetry $\Phi \rightarrow e^{i \theta} \, \Phi$, which the original Yang-Mills theory does not have. We will eliminate this unwanted symmetry by requiring that all physical operators must have equal numbers of 
$\Phi$'s and $\Phi^*$'s.

We will now show how this action leads to the wave function (\ref{12}), before discussing further properties.
The equations of motion corresponding to (\ref{21}) are
\beqar
- (D_\mu F_{\mu\nu} )^a ~+~ {\delta S_m \over \delta A_\nu^a} &=& e \left( (D_\nu \Phi)^\dagger
T^a \Phi - \Phi^\dagger T^a D_\nu \Phi \right)\label{22}\\
D_\mu \left( \sigma^\mu \, \sigma^\nu \, D_\nu \Phi \right) & =& 0\label{23}
\eeqar
We will use an expansion scheme to solve these equations in the following way.
In the first equation, the term linear in $A$ in the mass term will be kept
at the lowest order, but we will treat the effect of the current due to $\Phi$ (the right hand side of (\ref{22})), as well as nonlinear terms on the left hand side, in a perturbative expansion.
We will solve the second equation as it is.  
This expansion scheme is thus similar to what we did in the 
Hamiltonian approach in \cite{{KKN2}, {KNY}}.
This means that we can treat the Yang-Mills part and the $\Phi$-dependent terms of $\Gamma$ in
(\ref{21}) separately to the lowest order.
The quadratic term in $S_m (A)$, for any choice of the mass term, has the same form, namely,
$\sim A^{T2}$. Writing $A^T_\mu = A_\mu - \int_y \, \del_\mu G(x, y) \del\cdot A (y)$, we see that it is invariant under the (Abelian) gauge transformation $A_\mu \rightarrow A_\mu + \del_\mu \theta$, provided $G(x,y)$ obeys Dirichlet conditions and $\theta$ vanishes at $\tau =0, \beta$. 
(The use of Dirichlet conditions for the Green's functions is also emerges from an analysis of
BRST invariant boundary conditions \cite{moss}.) In this case, we can write
\beq
S_m (A) = {m^2 \over 2} \int A^{T2} ~+\cdots = {m^2 \over 2} \int \left[ A^2 - \del\cdot A (x) \, G(x,y) \, \del\cdot A (y) + \cdots \right]
\label{24}
\eeq
For the Yang-Mills part of the action, we then find
\beq
\delta W_{YM} = \int d^2x~ F^T_{0i} \delta A^T_i ~+~ \int d^2x~ {1\over 2} \left[ - F_{0i}^2 + A^T_i ( k^2 + m^2) A^T_i \right] \, \delta \beta
\label{25}
\eeq
This leads to
\beq
W_{YM} =  {1\over 2} \int d^2x~ A^T_i \sqrt{k^2 + m^2}\, A^T_i ~+\cdots
\label{26}
\eeq
upon setting ${\cal H}_E$ to zero.
This is entirely as expected when the gluon has a mass. In the $A_0 =0$ gauge, for the $\Phi$-dependent terms, we find
\beqar
\delta W &=& \int \left[ \delta \phi_1^\dagger ( {\dot \phi}_1 + 2\, {\bar D} \phi_2 ) + \delta \phi_2^\dagger
({\dot \phi}_2 - 2 \, D \phi_1) ~+ ~c.c.\right] ~+~ {\cal H}_E \, \delta \beta\nonumber\\
{\cal H}_E &=&\int \left[ 4 ({\bar D} \phi_2)^\dagger ({\bar D} \phi_2) + 4 ({\bar D} \phi_1)^\dagger
({\bar D} \phi_1) - {\dot \phi}_1 {\dot \phi}_1 - {\dot \phi}_2 {\dot \phi}_2 \right]
\label{27}
\eeqar
Solving ${\cal H}_E = 0$, we find $W_\Phi = \Phi^\dagger \, K \, \Phi$, with
\beq
K= 4\, \left[\begin{matrix}
0 & {\bar D} \\
- D & 0\\
\end{matrix}\right], \hskip .3in \Phi^a = \left( \begin{matrix}
\phi^a_1\\
\phi^a_2\\
\end{matrix}
\right)
\label{28}
\eeq
As I mentioned before, the field $\Phi^a_A$ is to be considered as an auxiliary field and observables are only made of
the Yang-Mills fields. For such an observable ${\cal O}$,
\beqar
\la {\cal O}\ra &=& \int d\mu (A) \, [d\Phi] \, ~~\Psi^*_{YM} \Psi_{YM}~~
\Psi^*_\Phi \, \Psi_\Phi ~~{\cal O} = \int d\mu (A) \, [d\Phi] \, ~~ \Psi^*_{YM} \Psi_{YM}~
e^{ - 2 W_\Phi}~~{\cal O}
\nonumber\\
&=&  \int d\mu (A) \, ~~\Psi^*_{YM} \Psi_{YM} \, ~~{1\over \det K } \, ~~{\cal O}
\sim \int d\mu (A) \, ~~\Psi^*_{YM} \Psi_{YM} \, ~~{1\over \det ( - D\bD )} \, ~~{\cal O}\nonumber\\
&\approx& \int d \mu (A) \, \Psi^*_{YM} \Psi_{YM} \, \,\exp\left( m \int \! A^{aT} A^{aT} + \cdots\right)\, \, {\cal O}\label{29}
\eeqar
where we have designated $e^{-W_{YM}}$ by $\Psi_{YM}$. This is equivalent to using
\beq
\Psi_0 \sim \exp \left[ - {1\over 2} \int A^{aT}_i (x) \left[{\sqrt{k^2 + m^2}~  { {- m}} }\right]_{x,y} ~A^{aT}_i (y) ~+\cdots\right]\label{30}
\eeq
where we used the result $\det ( - D \bD ) = \exp ( 2\, c_A \,S_{wzw}(H) )$.
This completes our demonstration that the effective action (\ref{21}) does indeed have the terms needed to obtain the wave function (\ref{10}) or (\ref{12}).

\section{Comments, Observations}

We can systematically improve upon the effective action by considering terms with higher powers of $J$ in the wave function. We have to work out terms to, say, cubic order in the $A$'s, starting from 
the action (\ref{21}), and compare them with the corresponding terms in the wave function.
If there is agreement, well and good. If not, we add a term to $\Gamma$ which has at least three powers of $A$ and match with the wave function to that order. In this way, we can systematically improve on the effective action.
However, we will not pursue this further here. The rest of this talk will be made of some comments on the
nature of $\Gamma$.
\begin{enumerate}
\item Is there a preferred choice of the gauge-invariant mass term?

There have been any different gauge-invariant mass terms which have been used in resummation schemes to estimate the gluon mass \cite{{mass}, {AN}}.
Our calculations so far do not show any preference for any particular one since  we have  only looked at the quadratic term. It has been noted that many of the mass terms have threshold singularities at $k^2 =0$, suggesting that there are still massless excitations left in the theory \cite{JP2}.
The one suggested in \cite{AN} does not have this feature. But this may not mean very much in the present context. It could be that with the AN mass term, we have a minimal form for the $\Phi$-dependent part of $\Gamma$. Some of the other mass terms presumably work just as well, but with a modification of the $\Phi$-dependent terms in (\ref{21}).
\item What can we say about excited states?

It is possible to discuss excited states as well, by a deformation of the wave function.
For example, for a simple scalar field theory, a deformation
$W \rightarrow W + \int j \, \phi$ corresponds to a modified Schr\"odinger equation of the form
\beq
( {\cal H} ~+~j{\rm -dependent ~ terms}) \, e^{-W - \int j \, \phi} = 0
\label{31}
\eeq
Expanding this in powers of $j$ will lead to wave functions for all excited states.
One could do this for two-body states as well. For example, by considering
$W \rightarrow W + \int_{x,y} j(\vx, \vy) \, \phi (\vx )\, \phi (\vy)$, we can get a two-body Schr\"odinger equation for the bound state of two particles. In the reduction of the higher powers of $\phi$ which can appear in such an equation, we will need to use partial Wick contractions. These will be determined by the vacuum wave function, so that ultimately, the interparticle potentials will be determined by the
vacuum wave function. For the gauge theory, we can consider a glueball state 
with $W \rightarrow W + \int f(\vx, \vy) F^a_{ij} (\vx) U^{ab}( \vx, \vy) \,F^b_{ij} (\vy)$.
(This is for a $0^{++}$ glueball.)
Since the interparticle potential arises from the vacuum wave function, and we already know that it leads to a nonzero string tension, we expect two-body equations of the form
\beq
\left[ \sqrt{m^2 - \nabla_1^2} \,+\, \sqrt{m^2 - \nabla_2^2} 
\, + \sigma_A \, \vert \vx - \vy\vert \right] f (\vx, \vy) \approx 0
\label{32}
\eeq
We expect that this can lead to a method to tackle the question of glueballs in our approach.

\item What are the implications of the auxiliary field $\Phi_A^a$?

In the gauge theory, we do expect some sort of gluon mass. But there is the compatibility between gluon masses (which would naively cut off any long range forces) and the existence of linear potentials, an issue which has been pointed out by Cornwall \cite{cornwall1} (and others). Our formula for the action (\ref{21}) shows that there is an additional crucial term in $\Gamma$, namely, the
$\vert \sigma\cdot D \Phi\vert^2$ term. This is presumably what allows us to have it both ways: gluon masses and long range forces.

Another important feature of $\Phi_A^a$ has to do with center vortices. As emphasized by many people, center vortices are important in understanding confinement for ${\mathbb Z}_N$-noninvariant representations and screening for ${\mathbb Z}_N$-invariant representations \cite{Zn}.
One can also argue for the existence of such vortices using
auxiliary fields for gluon masses \cite{cornwall1}. In our case, the $\Phi$-fields do have center vortices with well-behaved short distance properties. The absence of a potential term suggests that they are like BPS solutions. A quick analysis suggests also that the total vortex charge must be zero for reasons of Lorentz invariance. Thus multiple vortices of zero net charge are possible.
These could be the seed for understanding the screening versus confinement issue in our approach.

\end{enumerate}
\vskip .1in
I thank the organizers for inviting me to this very interesting workshop and for accommodating my difficult schedule. After this talk was presented at the workshop, some of the work has been written up as a research article \cite{nair2011}; it contains some more details and derivations.

\end{document}